\documentstyle[aps,epsfig,float,twocolumn,prl]{revtex} 
\newcommand{\be}{\begin{equation}}
\newcommand{\ee}{\end{equation}}

\draft
\begin{document}
\twocolumn[\hsize\textwidth\columnwidth\hsize\csname @twocolumnfalse\endcsname 

\title{Multiscaling at Point $J$: Jamming is a Critical Phenomenon}
\author{J. A. Drocco$^1$, M. B. Hastings$^{2,3}$, C. J. Olson Reichhardt$^3$, 
and C. Reichhardt$^{2,3}$}
\address{
$^1$Department of Physics, University of Notre Dame, Notre Dame, Indiana 46656\\
$^2$Center for Nonlinear Studies and $^3$Theoretical Division,
Los Alamos National
Laboratory, Los Alamos, New Mexico 87545}

\date{\today}
\maketitle
\begin{abstract}
We analyze the jamming transition that occurs as a function 
of increasing packing density 
in a disordered two-dimensional assembly of
disks 
at zero temperature for ``Point J'' of the recently proposed jamming
phase diagram.  
We measure the total number of moving disks and 
the transverse length of the moving region,
and find a power law divergence as the packing density increases toward
a critical jamming density. 
This provides evidence that
the T = 0 jamming transition 
as a function of packing density is a {\it second order} 
phase transition.
Additionally we find evidence for multiscaling, indicating the
importance of long tails in the velocity fluctuations. 
\end{abstract}
\vspace{-0.1in}
\pacs{PACS numbers: 45.70.-n, 64.60.Ht}
\vspace{-0.3in}

\vskip2pc] 
\narrowtext
There has been a surge of activity in jamming phenomena for 
$T=0$ systems such as granular materials, foams, and
colloids, where jamming is
defined as the onset of a nonvanishing yield stress in
a disordered state\cite{jr,Liu}. 
Liu and Nagel proposed a jamming phase diagram containing three
axes: temperature $T$, the inverse packing fraction $1/\phi$, 
and shear $\sigma$ \cite{jr}. The
system is jammed within a three-dimensional dome;
above the jamming transition, the system behaves
as a rigid solid. 
Recently O'Hern {\it et al.} \cite{j} studied the
area of the jamming phase diagram 
at $ T = \sigma = 0$ 
as a function of packing density $\phi$,
and 
showed that a well defined sharp jamming transition appears
at ``Point J.''
Elsewhere on the jamming phase diagram, the boundary between
jammed and unjammed states is not sharp since its definition is
sensitive to the experimental time scale.
It was noted in Ref.~\cite{j} that, although Point J does not exist
for liquids, the behavior near Point J 
could be relevant to the physics near the glass transition. 
A key question is whether Point J is a true
continuous phase
transition with a diverging spatial correlation length. 
Since the physics at Point J 
is non-thermal, any continuous phase transition behavior could be
dominated by rare fluctuations.
If this is the case, simple scaling close to the critical density 
may {\it not} occur even if there is an
underlying phase transition.  

Here, we consider the motion of a single
disk driven through an arrangement of disks
as the jamming transition is approached,
a system which, 
as suggested in Ref.~\cite{j}, 
should provide a direct test of 
the nature of the jamming transition. 
The reader is invited to try the following experiment:
place a large collection of coins flat on a desk, so that they are almost
touching.  Then, 
push one coin and observe what other coins move.
At low packing fractions 
$\phi$, the driven disk or coin occasionally contacts other disks, which can in
turn contact still other disks, but the total number of disks moving
is small.  At a certain density $\phi=\phi_c$,
however, the entire system must move simultaneously as a unit, and {\it jams}.
Force chains can be observed emanating from the driven disk \cite{fc}.

In these jamming problems, the tail of the
particle velocity distribution $P(v)$ at small velocities
controls the behavior of the system, since once the system enters
a jammed state, with everything stopped in the thermodynamic limit,
it can never exit the jammed state.  The best method for characterizing 
the tail of $P(v)$ is via multiscaling, which we employ
in this Letter.
For fixed driving force,
a small $v$ implies that a large number of other particles are dragged
with the driven particle, but we can also characterize the number of particles
moved by the driven particle by examining the force transmission.  We will see
that this measurement provides direct evidence for a diverging length
scale.
Our data also suggests that the jamming transition
for very large systems coincides with the random close packed density.

We simulate a binary mixture of two-dimensional disks 
with stiff spring repulsive interactions of radius $r_{A}$ 
and $r_{B}$ at $T = 0$
(see Fig.~1).
For all the densities 
we consider here, we find $\ll 1\%$ overlap in the radii 
as the disks interact, indicating that the spring constant is sufficiently
large to provide a good approximation to hard core disks.
The bimodal disk distribution, with a size ratio of $1.4:1$,
is chosen to create a disordered arrangement and
avoid formation of a regular lattice. 
We also performed
simulations with a size ratio of $1.7:1$ with substantially identical results.
We employ overdamped dynamics such that the velocity of each 
disk
is proportional to the force acting on it.  
The equation of motion for the disks is 
\be
\label{eqm}
\eta{\bf v}_{i}
= \sum_{i \neq j}{k}(|{\bf r}_{ij}| - r_{eff})
\frac{{\bf r}_{ij}}{|{\bf r}_{ij}|} + {\bf f}_{d} .
\ee
Here $k = 200$ is the strength of the stiff restoring spring \cite{MD},
${\bf r}_{ij} = {\bf r}_{i} - {\bf r}_{j}$, and for all but one
of the disks, the external driving force ${\bf f}_d=0$.
The single driven disk (large dot in Fig.~1) has ${\bf f}_d=0.1.$ 
Disks only interact if they are separated by
a distance smaller than the sum of their radii,
$|{\bf r}_{ij}| < r_{eff} = r^{(i)}+r^{(j)}$, where
$r^{(i)}$ equals either $r_A$ or $r_B$, depending on the given
particle.
We tested several different values
of ${\bf f}_{d}$ and chose the value small enough to minimize
overlap as mentioned above.
We use periodic boundary conditions with a range of linear system sizes $L=24$ 
to 60.  For 
$L=60$, the
system contains $N=2600$ background disks
at a density of $\phi=0.8395$.

\begin{figure}
\begin{center}
\epsfxsize=3.4in
\epsfbox{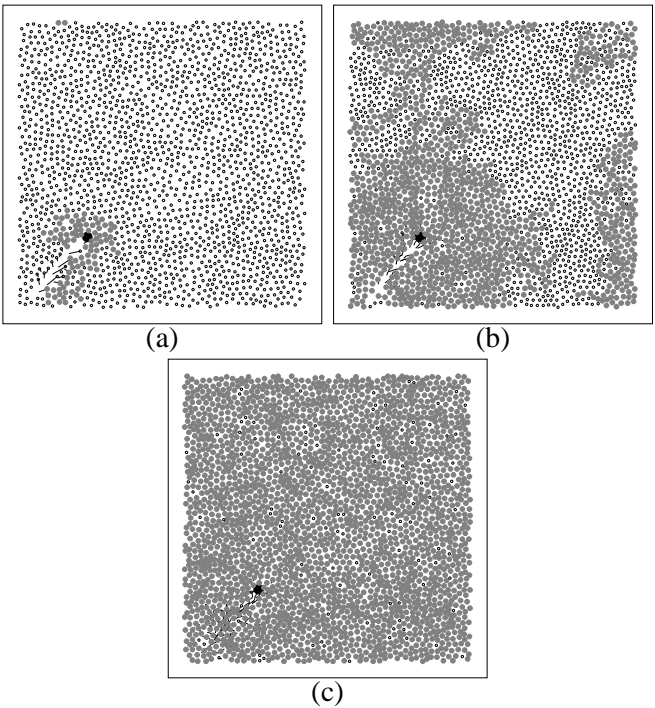}
\end{center}
\caption{Sample geometry for $L=60$ and (a) $\phi=0.656$,
(b) $\phi=0.811$, (c) $\phi=0.837$.
Large black dot: the probe disk (size exaggerated for clarity);
gray dots: moving disks; small dots: non-moving disks.
}
\end{figure}

We prepare the system by randomly dropping 
nonoverlapping disks up to $\phi\approx 0.6$.
To reach higher densities, we add disks at
randomly selected interstitial locations,  
reduce the radii of all disks, 
and then increase the radii back to the initial values while allowing the 
system to evolve under a small temperature.  This produces
nonoverlapping configurations at densities up to $\phi_c$.
To reach the maximum possible density,
$\sqrt{3}\pi/6\approx 0.907$, requires phase separating the system.
At
the significantly lower $\phi_c\approx 0.839$ we find no phase separation.
Since the system is not jammed below $\phi_c$, we believe that
we successfully uniformly sampled all available states,
and 
that phase separation does
not occur until some $\phi_{sep}>\phi_c$.

To study the velocity curve,
we drive a single large disk along a 45$^\circ$ angle 
over a distance of $\sqrt{2}L/5$.  
The time required for this motion is much longer
than the time scale of any brief initial transients in the velocity. 
We measure 
the effective velocity of the driven disk
over the length of the simulation. 
For computational reasons, it is not possible to simulate 
a disk
moving at an arbitrarily slow velocity through the system.  In these
cases, we declared that the system had jammed; this definition
agreed well with the critical $\phi_c\approx 0.839$ given by 
curve fitting below.

In Fig.~1 we show snapshots of simulation results 
for different densities, with  
disks counted as moving and shaded in gray
if they were connected via a force
contact to the driven disk.
At the lowest 
$\phi=0.656$, 
in Fig.~1(a),
the driven disk moves easily and interacts 
with 
only a 

\begin{figure}
\center{
\epsfxsize=3.4in
\epsfbox{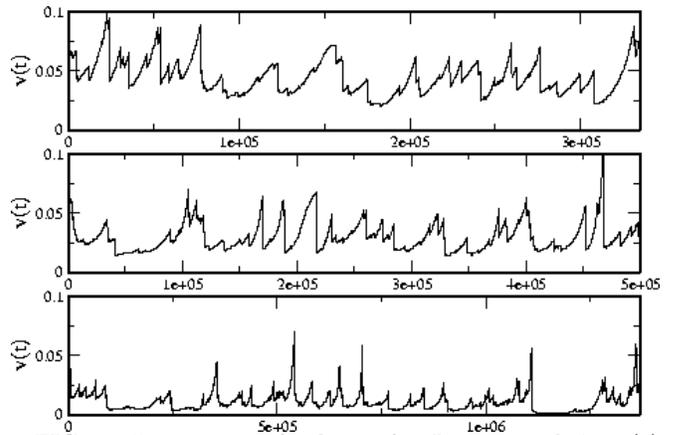}}
\caption{Time series of velocity for $L=60$ and $\phi=$ (a) 0.656, (b) 0.747,
and (c) 0.837.}
\end{figure}

\noindent
few neighbors. 
In Fig.~1(b) for 
$\phi = 0.811$, 
close to jamming, a larger portion of the disks are moving,
while in Fig.~1(c) at $\phi = 0.837$, the system is
sufficiently close to $\phi_c$ that the entire (finite-size) sample is moving.

Figure 2 shows example time series 
for the velocity $v$ parallel to the applied drive
at $\phi=0.656$, 0.747, and 0.837.
Not only does the average 
velocity 
decrease as $\phi$ increases,
but the time series becomes more intermittent. At $\phi=0.837$, 
the velocity 
is usually very small, but there are
occasional bursts of much higher 
velocity.

We emphasize that
for $\phi<\phi_c$
the threshold force 
required to move the driven particle vanishes.  
Instead, 
$v$ is proportional to the drive $f_d$ at any moment of time.  This
follows from dimensional grounds, since for a hard-core system, there is
no physical parameter with units of force.  
This is very different
from the behavior shown in
a recent numerical study of single probe particles driven through a 
system with softer, long-range forces \cite{softglass},
where there is a non-vanishing, but finite,
depinning force 
at all densities due to the long-range nature of the force.

{\it Distribution of Velocities---}
Newton's third law, combined with Eq.~(\ref{eqm}), leads to the
result that $\eta \sum_i {\bf v}_i={\bf f}_d$.  Thus, if the driven
particle is not in contact with any other particles, it moves at 
$v=f_d/\eta$.  If there are a total of $n$ particles moving together,
including the driven particle, then they each move at
$v=f_d/n$.  
For $\phi>\phi_c$, all $N$ particles in the system 
move
together, and 
$v$ is vanishingly small in the thermodynamic limit.
For $\phi<\phi_c$, the number of 
particles moving is finite in the thermodynamic
limit and diverges as $\phi$ approaches $\phi_c$.

One measurement which may
show scaling as $\phi$ approaches $\phi_c$ is the average velocity 
$v$ of the
driven disk.  However,
$v$ is a random, time-dependent
variable, and
for any density $\phi<\phi_c$ there is a nonvanishing probability of observing
any given, arbitrarily small 
$v$.
Since the other disks are distributed randomly, there is still some 
probability that in any local region the density will 
exceed $\phi_c$ \cite{lt} (this argument is inspired by Lifshitz
tails).  If this region 
encloses $n$ disks, 
the 
probability of finding such a 

\begin{figure}
\begin{center}
\epsfxsize=3.4in
\epsfbox{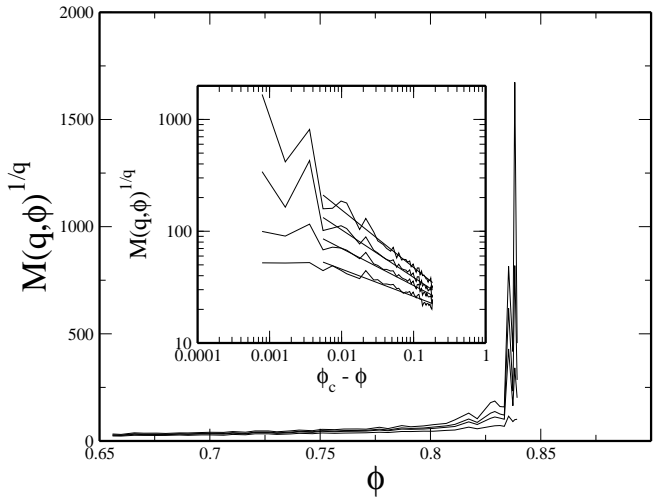}
\end{center}
\caption{Plot of $M(q,\phi)^{1/q}$ versus $\phi$ for $q=-1$, 1, 2, and 4 
(from bottom to top).  The curves with larger $q$
are noisier.  Inset: $M(q,\phi)^{1/q}$ versus $\phi_c-\phi$ on
a log-log scale for $q=-3,-1,1,3$. A curve fit shows the power law scaling. }
\end{figure}

\noindent
region is exponentially
small in $n$, but it is nonvanishing.  When the driven disk
hits such a region, it behaves as if it jams, and  
$v$ slows to
a value of order $1/n$ until the disk either escapes the region or 
pushes
other disks 
into neighboring
regions of lower density.  Thus, with exponentially small probability in
$n$, $v \sim O(1/n)$.  Once the particle slows, however, it takes a long
time for it to leave the region.  Thus, one expects to see
an intermittent 
$v$.

To measure the intermittency, we use multifractal\cite{mf} scaling.
For a given $\phi$, let
$p(v) {\rm d}v$ be the probability of measuring a given 
$v$
at a single instant in time.  Define the $q$-th inverse moment by
$M(q)=\int {\rm d}v p(v) v^{-q}$ as the time average of the $q$-th power of the
inverse
velocity.  We then define a set of multifractal exponents
$\tau(q)$ by
\be
M(q)=
\int {\rm d}v \, p(v) v^{-q} \propto (\phi_c-\phi)^{-\tau(q)}.
\ee
If 
$v$
were constant throughout the simulation for a given $\phi$, we would
have $M(q)^{1/q}=M(1)$.
If instead 
$v$ had relative
fluctuations of order unity about some characteristic velocity, $v_0(\phi)$,
then $M(q)^{1/q}$ would 
not equal $M(1)$ in general, but
$\tau(q)/q$ would still be independent of $q$.
We will instead find that
$\tau(q)/q$ {\it depends} on $q$.  This implies that 
$v$ does
not simply fluctuate about a single 
$v_0$, but is instead much more intermittent, 
with the particle
sometimes getting stuck for a long time at a slow velocity, then traveling
much more rapidly.

We compute the moments $M$ and extract the exponents $\tau(q)$
by fitting the moments to the form $c*(\phi_c-\phi)^{-\tau(q)}$.  
The
value of $\phi_c$ determined from these fits is 
independent
of $q$ to high accuracy for $q\geq -2$, and thus we can fix 
$\phi_c\approx 0.839$ as the onset of jamming.  
In Fig.~3 we plot $M(q,\phi)^{1/q}$ versus $\phi$ for $L=60$ and
various $q$, averaged over three realizations of the system.
In the inset we show the log-log plot of the curves
in the main panel 

\begin{figure}
\begin{center}
\epsfxsize=3.4in
\epsfbox{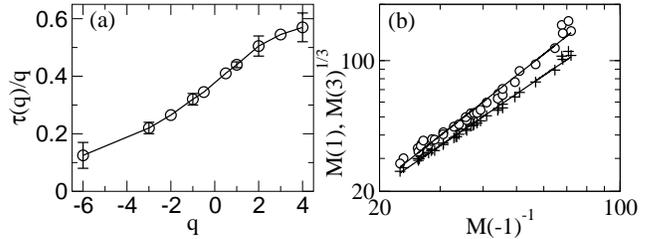}
\end{center}
\caption{(a): Plot of $\tau(q)/q$ against $q$.  The $q$ dependence of
$\tau(q)/q$ shows the existence of multiscaling in this system.
(b): Plot of $M(1)$ (crosses) and $M(3)^{1/3}$ (circles) against $M(-1)^{-1}$.}
\end{figure}

\noindent
with solid lines indicating power law fits.  The
curves not only
fail to overlap, but also
have different slopes, indicating the presence of multiscaling.
In Fig.~4(a), we plot $\tau(q)/q$ versus $q$.  For large positive
and negative $q$, the plot asymptotes.  The asymptote at large positive $q$
reflects the scaling of the typical (disregarding exponentially
rare possibilities discussed above) slowest velocity
of the system which goes to zero as $(\phi-\phi_c)^{\lim_{q\rightarrow\infty}
\tau(q)/q}$.  Similarly, the asymptote at negative $q$ reflects the typical
largest velocity.  Since $\lim_{q\rightarrow -\infty}\tau(q)/q$ is very close
to zero, the typical largest velocity is largely independent of $\phi$, as
seen in Fig.~2.  In fact, it is likely that at sufficiently negative $q$ the
exponent $\tau(q)/q$ becomes zero.

The error in $\tau(q)$ due to statistical fluctuations is negligible.
If we compare the $\tau(q)$ obtained by curve fitting $M(q,\phi)$ averaged
over all realizations to $\tau(q)$ obtained by using only a single 
realization, the
difference in exponents is of order $0.001$ to $0.002$.  Instead,
the major 
source of error is finite size effects.  The straight lines in the
inset to Fig.~3 are based on fitting over a certain range of $\phi$.  For
$\phi$ closer to $\phi_c$ than the endpoint of the 
fitting lines, the statistical
noise in the data increases significantly and some of the realizations
jam while others do not.
Thus, for finite $N$ the
jamming threshold is not well defined, and these $\phi$ are so close to
$\phi_c$ that finite size effects may be important.  The error bars in
Fig.~4 are based on this consideration of finite size effects.  The low end
of the error bars corresponds to the $\tau(q)$ obtained by fitting only
over the range shown in the inset to Fig.~3, while the high end corresponds
to including all $\phi<\phi_c$.  The low end tends to underestimate the
difference in $\tau(q)/q$ for different 
$q$.  $\tau(q)/q$ is clearly $q$-dependent since the difference in
$\tau(q)/q$ between $q=-1$ and $q=2$, for example, 
is definitely outside
the error bars.
In Fig.~4(b) we plot $M(q)^{1/q}$ as
a function of $M(-1)^{-1}$ for $q=1,3$ over the 
same range as in
Fig.~3 to illustrate that the data obey extended self-similarity \cite{ess}.  
The slopes on the log-log plot are
$1.25\pm 0.15$ and $1.45 \pm 0.2$ 
respectively; as expected, this is within error bars of
the ratio of the exponents $-\tau(1)/\tau(-1)$ and $-(\tau(3)/3)/\tau(-1)$.
It is difficult to establish that these slopes are different from
each other, but the slopes are definitely different from unity, which
already indicates multiscaling.
The fact that $\tau(-1)$ is 

\begin{figure}
\begin{center}
\epsfxsize=3.4in
\epsfbox{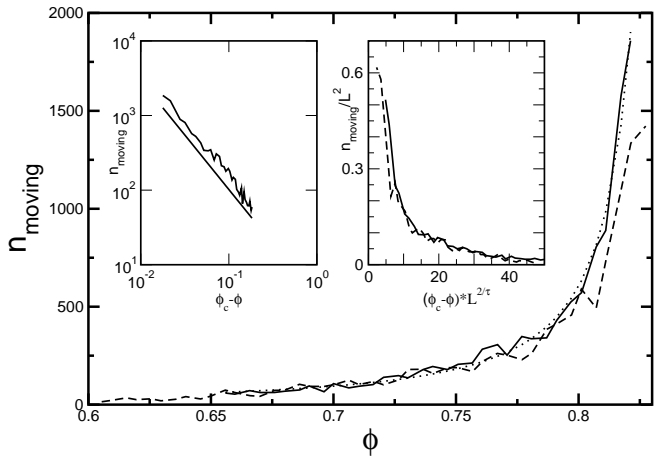}
\end{center}
\caption{Plot of $n_{\rm moving}$ against $\phi$.  Solid line: $L=60$.
Dashed line: $L=48$.  Dotted line: a power law fit to the $L=60$ curve.
Left inset: The $L=60$ curve and the fit against $\phi_c-\phi$,
showing scaling. Right inset: finite size scaling.}
\end{figure}

\noindent
less
than one makes the size of the  scaling regime seem small, since a wide
scaling range in $\phi$ leads to only a small range in $M(-1)$;
however, the curve fit is very good over the available range.

Figure 1 shows that as the jamming transition is approached, 
the number of moving disks increases and there is a diverging
length scale as the jamming is approached. In order to quantify this,
we show in Fig.~5 
the number of moving disks $n_{\rm moving}$
vs $\phi$ for systems 
of linear size 
$L=48$ and $L=60$. 
Here a clear divergence appears as the 
critical density is approached.  The divergence is cut off when
$n_{\rm moving}$ equals the total number of disks.
We have also considered smaller systems 
for different 
parameters 
of drive and disk radii and again 
observe a divergence; however, these smaller systems give
a much lower resolution and hence a larger error on the estimated exponent.  
In Fig.~5 we fit a power law to the largest 
system with $(\phi_c - \phi)^{\tau}$, where $\tau$ is between $1.2$ and $1.46$.
We also measured the sum over moving disks of the squared distance
between the driven disk and the moving disk, 
a quantity $n_{\rm moving}l^2$, and find that this number
diverges as $(\phi_c-\phi)^{-\sigma}$, with $\sigma$ between 2.35 and 2.6.
The number of moving disks cannot directly be compared to the
velocity $v$ of the driven disk, as
the driven disk can push other disks
normal to the drive so that $n_{\rm moving}$ may be much larger than $1/v$.
We obtained the exponents $\tau$ and $\sigma$ by a time average of the
number of moving disks, and did not perform a
multiscaling calculation, though we can extract a length
from a simple scaling analysis.  If the moving disks form a cluster
with dimension $d$ and length scale $\xi$ diverging as 
$\xi\propto (\phi_c-\phi)^{-\nu}$, then $\tau=\nu d$ and $\sigma=\nu (d+2)$.
Given the values of $\tau$, $\sigma$, and our own qualitative observations,
we find $d=2$, so that $\nu$ is between 0.6 and 0.7.  This provides
a direct measurement of the diverging length scale, in reasonable agreement
with the finite-size scaling result \cite{j} $\nu = 0.71 \pm 0.12.$
We note that Fig.~5 in our case is completely consistent with finite-size
scaling as shown in the right inset.
Scale the $y$-axis by $L^2$ (the scaling for
the total number of particles in the system), and scale $\phi_c-\phi$ by
$L^{-2/\tau}$.  Then, the curves for different $L$ automatically
collapse within the scaling region, and since the divergence is
cutoff at $n_{\rm moving}\propto L^2$ for all $L$, these curves also
collapse close to $\phi_c$.
Finally, our
$\phi_{c} \approx 0.839$ is very close to the critical packing density found in \cite{j}. 

{\it Conclusion---}
We have studied a system
of a $T=0$, zero shear 2D disordered assembly of disks
at densities below and up to  
Point J in the recently proposed jamming phase diagram.
A single probe disk is pushed with a constant drive through
the other disks. Upon increasing the packing density,
we find a jamming transition associated with a power
law divergence in the number of moving disks
and a diverging spatial correlation 
length, indicating that 
Point J is a true
continuous phase transition.
In addition, we
show
that the tails of the disk velocity distribution 
play an important role in this 
transition, since due to the non-thermal nature of the
system, once the particle stops moving,
it cannot restart.
Using the multifractal moments $\tau(q)$, we chart the tails 
and show that the $\tau(q)$ do not depend linearly on $q$;
thus, the system exhibits multiscaling.  
We also find evidence for a diverging correlation length
of the force contacts
as the jamming transition is approached from below.
Our results 
show that there is an underlying 
{\it second order} non-thermal phase transition
at the  
jamming transition. 
We suggest that experimental
work should consider both the velocity time series and
the number of moving particles in the surrounding media,
and should test for 
multiscaling behavior
as the jamming transition is approached.
Particular experiments would include driving a single disk on a
flat surface through a disordered assembly of other disks for increasing
density, or driving individual colloids through glassy assemblies of 
other colloids. 

{\it Acknowledgments---}
We thank E. Weeks for useful discussions.
This work was supported by the U.S. DOE under Contract No.
W-7405-ENG-36.

\end{document}